\documentclass[runningheads]{llncs}
\usepackage[T1]{fontenc}   
\usepackage{amsmath}
\usepackage{graphicx}
\newtheorem{fact}{Fact}
\newcommand{\junk}[1]{}

\begin{document}
%\documentclass{cccg23}
%\usepackage{graphicx,amssymb,amsmath}

%----------------------- Macros and Definitions --------------------------

% Add all additional macros here, do NOT include any additional files.

% The environments theorem (Theorem), invar (Invariant), lemma (Lemma),
% cor (Corollary), obs (Observation), conj (Conjecture), prop
% (Proposition), and proof are already defined in the cccg19.cls file.
% Add additional environments only if you REALLY need them.

%----------------------- Title -------------------------------------------

\title{Convex Hulls, Triangulations, and Voronoi Diagrams
  of Planar Point Sets on the Congested Clique\thanks{A
preliminary version of this article appeared in
\emph{Proceedings of the Thirty-Fifth Canadian Conference on Computational
Geometry} (CCCG~2023), Montr\'{e}al, Canada, pp.~183--189, 2023.}}
\author{Jesper Jansson
  \inst{1}
  \and
  Christos Levcopoulos
  \inst{2}
  \and \\
  Andrzej Lingas
  \inst{2}
\and
  Valentin Polishchuk
  \inst{3}
}
%\authorrunning{J. Jansson, C. Levcopoulos, and A. Lingas}
\institute{
Graduate School of Informatics, Kyoto University, Kyoto, Japan.
\email{jj@i.kyoto-u.ac.jp}
\and
Department of Computer Science, Lund University, Lund, Sweden. 
\email{$\{$Christos.Levcopoulos, Andrzej.Lingas$\}$@cs.lth.se}
\and
  Communications and Transport Systems, ITN, Link\"{o}ping university, Sweden
\email{valentin.polishchuk@alumni.stonybrook.edu}}

\pagestyle{plain}
\maketitle
%%%%%%%%%%%%%%%%%%%%%%%%%%%%%%%%%%%%%%%%%%%%%%%%%%%%%%%%%%%%%%%%%%%

\begin{abstract}
  We consider geometric problems on planar $n^2$-point sets in the
  congested clique model.  Initially, each node in the $n$-clique
  network holds a batch of $n$ distinct points in the Euclidean plane
  given by $O(\log n)$-bit coordinates.  In each round, each node can
  send a distinct $O(\log n)$-bit message to each other node in the
  clique and perform unlimited local computations.  We show that the
  convex hull of the input $n^2$-point set can be constructed in
  $O(\min\{ h,\log n\})$ rounds, where $h$ is the size of the hull, on
  the congested clique. We also show that a triangulation of the input
  $n^2$-point set can be constructed in $O(\log^2n)$ rounds on the
  congested clique. Finally, we demonstrate that the Voronoi diagram
  of $n^2$ points with $O(\log n)$-bit coordinates drawn uniformly at
  random from a unit square can be computed within the square with
  high probability in $O(1)$ rounds on the congested clique.
  \end{abstract}
%\end{frontmatter}
\section{Introduction}
%The \emph{congested clique} model was introduced by Lotker \emph{et al.}
%\cite{LP-SPP05} as a model of communication/computation that focuses on the
%cost of communication between nodes in a network and ignores that of local
%computation within each node.
%The communication/computation model of \emph{congested clique} focuses on the
%communication cost and ignores that of local computation.
%It can be seen as
%The \emph{congested clique} is a model of communication/computation that
%focuses on the cost of communication between the nodes in a network and ignores
%that of local computation within each node.
The \emph{congested clique} is a model of communication/computation that
focuses on the cost of communication between the nodes in a network and ignores
that of local computation within each node.
This model was introduced by Lotker \emph{et al.} \cite{LP-SPP05}.
It can be seen as a reaction to the criticized Parallel Random Access Machine
(PRAM) model, studied extensively in the 1980s and 1990s, which in contrast
focuses on the computation cost and ignores the communication cost
\cite{ACG85}.

\begin{figure}
%[h]
\begin{center}
\includegraphics[scale=0.7]{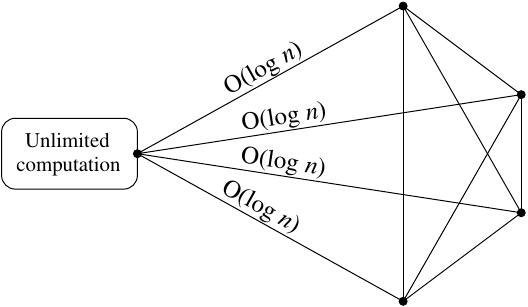}
\end{center}
\caption{An example of a congested clique network.}
\label{fig: computation1}
\end{figure}

In recent decades, the complexity of dense graph problems
has been intensively studied in the congested clique
model. Typically, each node of the clique network initially
represents a distinct vertex of the input graph and knows that
vertex's neighborhood in the input graph.
The nodes are assumed to have unique numbers (IDs) between $1$ and $n$ which are
already
known by all nodes in the network at the start of the computation.
Then, in each round,
each of the $n$ nodes can send a distinct message of $O(\log n)$ bits
to each other node and can perform unlimited local computation; see
Fig. \ref{fig: computation1}.
Several dense graph problems, for example, the minimum spanning tree
problem, have been shown to admit $O(1)$-round algorithms in the
congested clique model \cite{K_21,R}.
Note that when the input graph is of bounded degree,
each node can send its whole information to a distinguished node in
$O(1)$ rounds. The distinguished node can then solve the graph problem
locally. However, when the input graph is dense such a trivial solution
requires $\Omega(n)$ rounds.

Researchers have also
%successfully
studied problems not falling in the
category of graph problems, like matrix multiplication \cite{CK15} or
sorting and routing \cite{L}, in the congested clique model.
In both cases,
one assumes that the basic items, i.e., matrix entries or keys,
respectively, have $O(\log n)$ bit representations and that each node
% has a batch of $n$ such items initially. %%%
initially has a batch of $n$ such items. %%%
As in the graph case,
%it is assumed that %%%
%each node in each round can send a distinct $O(\log n)$-bit %%%
%message to each other node and perform unlimited computation. %%%
each node can send a distinct $O(\log n)$-bit message to %%%
each other node and perform unlimited computation %%%
in every round. %%%
Significantly, it has been shown that matrix multiplication
admits  an $O(n^{1-2/\omega})$-round algorithm \cite{CK15},
where $\omega$ is the exponent of fast matrix
multiplication,
while sorting and routing admit $O(1)$-round algorithms
(Theorems 4.5 and 3.7 in \cite{L})
%(Facts 1, 2)
under the aforementioned assumptions.
%while sorting and routing admit
%$O(1)$-round algorithms \cite{L} under the aforementioned assumptions.

We extend
this approach to include
basic geometric
problems on planar point sets. These problems are generally known to admit
polylogarithmic time solutions on PRAMs with a polynomial number
of processors \cite{ACG85}.
Initially, each node of the $n$-clique network holds a batch of $n$
points belonging to the input set $S$ of $n^2$ points with $O(\log
n)$-bit coordinates in the Euclidean plane. As in the graph, matrix,
sorting, and routing cases, in each round, each node can send a
distinct $O(\log n)$-bit message to each other node and perform
unlimited local computations. Analogously, trivial solutions
consisting in gathering the whole data in a distinguished node
require $\Omega(n)$ rounds.

More precisely, the problems that we consider are computing the convex hull,
a triangulation, and the Voronoi diagram of a set~$S$ of $n^2$
points with $O(\log n)$-bit coordinates in the plane,
defined next.
The \emph{convex hull of~$S$} is the smallest convex polygon~$P$ for which
every $q \in S$ lies in the interior of~$P$ or on the boundary of~$P$.
A \emph{triangulation of~$S$} is a maximal set of non-crossing edges between
pairs of points from~$S$.
Finally, the \emph{Voronoi diagram of~$S$} is the partition of the plane into
$|S|$~regions such that each region consists of all points in the plane having
the same closest point in~$S$.

Our contributions are as follows.
First, we provide a simple implementation of the Quick Convex Hull
algorithm \cite{E77}, showing that the convex hull of $S$ can be
constructed in $O(h)$ rounds on the congested clique, where $h$ is the size
of the hull. Then, we present and analyze a more refined algorithm
for the convex hull of $S$ on the congested clique running in $O(\log n)$
rounds. Next, we present a divide-and-conquer method for constructing
a triangulation of $S$ in $O(\log^2 n)$ rounds on the congested clique.
We conclude with with remarks on the construction of
the Voronoi diagram of a planar point set.
In particular, we show that
  the Voronoi diagram of $n^2$ points with $O(\log n)$-bit coordinates
       drawn uniformly at random from a unit square
       can be computed within the square with high probability
       in $O(1)$ rounds on the congested clique.
   
We also refer to the points of the input point set as \emph{vertices}, while
reserving the word \emph{nodes} to refer to the communicating parties in the
underlying congested clique network.
In order to simplify
  the presentation, we assume throughout the paper that the points in the input point sets
  are in general position.
  \junk{In the next section, we provide basic notation
  and definitions. In Section 3, we present our implementation of the Quick
  Convex Hull algorithm on the congested clique.
  In Section 4, we present and analyze our refined algorithm for
  the convex hull of a planar point set on the congested clique.
  In Section 5, we show our algorithm for a triangulation of
  a planar point set on the congested clique.
  In Final remarks, we discuss the construction of the Voronoi diagram
  of a planar point set on the congested clique.}

\section{Preliminaries}
%For a positive integer $r$, $[r]$ stands for the set of positive integers not exceeding $r.$

Let $S=\{ p_1,...,p_n\}$ be a set of $n$ distinct points in the Euclidean plane
such that the $x$-coordinate of each point is not smaller than that of $p_1$
and not greater than that of $p_n$.
%\footnote{In order to simplify
%  the presentation, we assume throughout the paper that the points in the input point sets
%  are in general position.}.
The {\em upper hull} of $S$ (with respect to $(p_1,p_n)$)
is the part of the convex hull of $S$ beginning in $p_1$ and ending in $p_n$
in clockwise order. Symmetrically, the {\em lower hull} of $S$ (with respect to $(p_1,p_n)$)
is the part of the convex hull of $S$ beginning in $p_n$ and ending in $p_1$
in clockwise order.
% %%%

\begin{figure}
%[h]
\begin{center}
\includegraphics[scale=0.7]{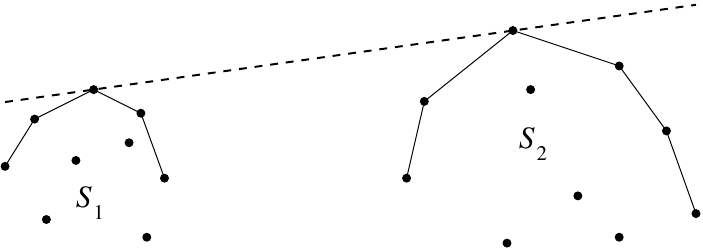}
\end{center}
\caption{An example of the bridge between the upper hulls of $S_1$ and $S_2.$}
\label{fig: bridge}
  \end{figure}

A {\em supporting line} for the convex hull or upper hull or lower hull
of a finite point set in the Euclidean plane is a straight line that touches the
hull without crossing it properly.
Let $S_1,$ $S_2$ be two finite sets of points in the Euclidean plane
separated by a vertical line.
The {\em bridge} between the upper (or lower) hull of $S_1$
and the upper (or, lower, respectively) hull of $S_2$ is
a straight line that is a supporting line for both of the upper (lower, respectively)
hulls.
See Fig.~\ref{fig: bridge} for an illustration. %%%

We define the {\em Information Distribution Task} (IDT) \cite{L}
as follows:\\
%Each node of the congested $n$-clique holds a set of 
%exactly $n$ $O(\log n)$-bit messages with their destinations.
Each node of the congested $n$-clique holds a set of exactly $n$
$O(\log n)$-bit messages with their destinations, with multiple messages from
the same source node to the same destination node allowed.
Initially, the destination of each message is known only to its source node.
Each node is the
destination of exactly $n$ of the aforementioned messages. The
messages are globally lexicographically ordered by their source node, their
destination, and their number within the source node.  For simplicity,
each such message explicitly contains these values, in particular
making them distinguishable. The goal is to deliver all messages to their
destinations, minimizing the total number of rounds.

Lenzen showed that   IDT can be solved in 16 rounds (Theorem 3.7 in
\cite{L}). He also observed that the relaxed IDT, where each node is required
to send and receive at most $n$ messages, easily reduces to IDT
in $O(1)$ rounds. Hence, we have the following fact.

\junk{Lenzen considered the {\em Information Distribution Task} (IDT)
\cite{L} which differs from GIDT in the requirement that each node
holds exactly $n$ messages and needs to receive exactly $n$ messages.
He showed  that IDT can be solved in 16 rounds (Theorem 3.7 in
\cite{L}). By using his result, each node can learn first about the
distribution of the node destinations of all messages so if necessary can
fill up the set of own messages with dummy messages with appropriated
destinations, in order to reduce GIDT to IDT in $O(1)$ rounds.  Hence,
by a multiple application of Theorem 3.7 in \cite{L}, we obtain the
following fact.}

\begin{fact}\cite{L}\label{fact: routing}
  The relaxed Information Distribution
  Task can be solved deterministically within $O(1)$ rounds.
  \end{fact}

The {\em Sorting Problem} (SP) is defined as follows:\\
Each node $i$ of the congested $n$-clique holds a set of $n$
$O(\log n)$-bit keys. All the keys are different w.l.o.g.  Each node
$i$ needs to learn all the keys of indices in $[n(i-1)+1,ni]$ (if any)
in the total order of all keys.

Lenzen showed that SP can be solved in 37 rounds if each node holds a
set of exactly $n$ keys (Theorem 4.5 in \cite{L}).  In order to relax
the requirement that each node holds exactly $n$ keys to that of with
most $n$ keys, we can determine the maximum key and add appropriate
different dummy keys in $O(1)$ rounds.  Hence, we obtain the following
fact.

  \begin{fact}\cite{L}\label{fact: sorting}
    The relaxed Sorting Problem can be solved in $O(1)$ rounds.
  \end{fact}

\section{Quick Convex Hull Algorithm on Congested Clique}
\label{section: quick}
The Quick Convex Hull Algorithm (also known as QuickHull or CONVEX) is well known in the literature;  see, e.g.,
\cite{E77,RS}. Roughly, we shall implement it as follows in the congested clique
model. First, the set $S$ of $n^2$ input points with $O(\log n)$-bit
coordinates is sorted by their $x$-coordinates \cite{L}.
As a result, each consecutive
clique node gets a consecutive $n$-point fragment of the sorted $S.$
Next, each node informs all other nodes about its two extreme points
along the $x$  axis. By using this information, each node can determine
the same pair of extreme points $p_{min}$, $p_{max}$ in  $S$ along the $x$ axis.
Using this extreme pair, each node can decompose its subsequence of $S$
into the upper-hull subsequence consisting of
the points that lie above or on the segment $(p_{min},p_{max})$ and
the lower-hull subsequence consisting of points that lie below or on $(p_{min},p_{max}).$
From now on, the upper hull of $S$ and the lower hull of $S$ are computed
separately by calling the procedures $QuickUpperHull(p_{min},p_{max})$ and $QuickLowerHull(p_{min},p_{max}),$
respectively.
The former procedure proceeds as follows.
Each node selects a point $q$ at maximum distance from the segment $(p_{min},p_{max})$
among all points
%those
in its upper-hull subsequence, excluding the points
$p_{min}$ and $p_{max}$. Next, it sends the point  $q$  to all other nodes.
\junk{Then, each node selects the
same point $q$ at maximum distance from the  segment $(p_{min},p_{max})$
among all points in the whole upper-hull
subsequence different
from $p_{min}$ and $p_{max}$}
Then, each node selects the same point $q$, different from $p_{min}$ and
$p_{max}$, at maximum distance from the segment $(p_{min},p_{max})$ among all
points in the whole upper-hull subsequence; see Fig. \ref{fig: pqr}. Note that $q$
must be a vertex of the upper hull of $S.$
Two recursive calls $QuickUpperHull(p_{min},q)$ and $QuickUpperHull(q,p_{max})$ follow.
  The procedure $QuickLowerHull$ is defined symmetrically.
  As each non-leaf call of these two procedures results in a new vertex of the convex hull,
  and each step of these procedures but for the recursive calls takes $O(1)$ rounds,
  the total number of rounds necessary to implement the outlined variant of
  Quick Convex Hull algorithm, specified in the procedure $QuickConvexHull(S),$
  is proportional to the size of the convex hull of $S.$
   \begin{figure}
%[h]
\begin{center}
\includegraphics[scale=0.7]{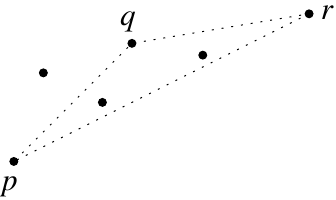}
\end{center}
\caption{Illustrating the points $p,\ q,\ r$
in the procedure $QuickUpperHull.$}
\label{fig: pqr}
  \end{figure}
    \par
    %\newpage
    \noindent
        {\bf procedure} $QuickConvexHull(S)$
     \par
        \noindent
            {\em Input}: A set of $n^2$ points in the Euclidean plane
            with $O(\log n)$ bit coordinates, each node holds a batch of
            $n$ input points.
            \par
            \noindent
                {\em Output}: The vertices of the convex hull of $S$
                held in clockwise order in consecutive nodes
                in batches of at most $n$ vertices.

    \begin{enumerate}
    \item Sort the points in $S$ by their $x$-coordinates so
      each node receives a subsequence consisting of $n$ consecutive
      points in $S$, in the sorted order.
    \item
    Each node  sends the first point and the last point in its subsequence
    to the other nodes.
    \newpage
    \item
      Each node computes the same point $p_{max}$ of the maximum $x$-coordinate
      and the same point $p_{min}$ of the minimum $x$-coordinate in the whole
      input sequence $S$
% on the base of %%%
based on %%%
the gathered information. (If there are ties in the minimum $x$-coordinate
then $p_{min}$ is set to a point with minimum $y$-coordinate.
Similarly, if there are ties in the maximum $x$-coordinate
then $p_{max}$ is set to  a point with maximum $y$-coordinate.)
\item
Each node
      decomposes its sorted subsequence into the upper hull subsequence
      consisting of points above or on the segment connecting
      $p_{max}$ and $p_{min}$ and the lower hull subsequence
      consisting of the points lying below or on this segment.
      In particular, the points $p_{min}$ and $p_{max}$
      are assigned to both upper and lower hull subsequences of the
      subsequences they belong to.
    \item
      Each node sends its first and last point in its upper hull subsequence
      as well as its first and last point in its lower hull subsequence
      to all other nodes.
    \item
      $QuickUpperHull(p_{min}, p_{max})$
    \item
      $QuickLowerHull(p_{min}, p_{max})$
    \item
By the previous steps, each node keeps consecutive
  pieces (if any) of the upper hull as well as the lower hull.
  However, some nodes can keep empty pieces.
  In order to obtain a more compact output representation
  in batches of $n$ consecutive vertices of the hull (but for the last batch)
    assigned to consecutive nodes of the clique, the nodes can count the number
    of vertices on the upper and lower hull they hold and send
    the information to the other nodes. Using the global
    information, they can design destination addresses for
    their vertices on both hulls. Then, the routing protocol
    from
    %\cite{L}
    Fact 1 can be applied.
    \end{enumerate}
    \par
    \vskip 5pt
   \junk{ \begin{figure}
%[h]
\begin{center}
\includegraphics[scale=0.7]{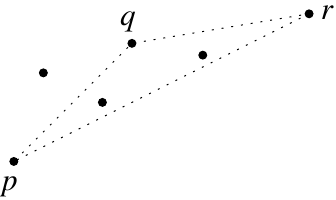}
\end{center}
\caption{Illustrating the points $p,\ q,\ r$
in the procedure $QuickUpperHull.$}
\label{fig: pqr}
  \end{figure}}
    \noindent
    {\bf procedure} $QuickUpperHull(p,r)$
    \par
    \noindent
    \par
        \noindent
            {\em Input}: The upper-hull subsequence
            of the input point set $S$ held in consecutive
            nodes in batches of at most $n$ points
            and two distinguished points $p,\ r$ in
            the subsequence , where the $x$-coordinate of $p$ is smaller
            than that of $r.$
            \par
            \noindent
                {\em Output}: The vertices of the upper hull of $S$
                with $x$-coordinates between those of $p$ and $r$
                held in clockwise order in consecutive nodes, between
                those holding $p$ and $r$ respectively, in batches
                of at most $n$ points.
    \begin{enumerate}
  \item
    Each node $u$ determines 
    the set $S_u$ of points in its upper-hull
    subsequence that have $x$-coordinates between
    those of $p$ and $r$ and lie above or on the segment
    between $p$ and $r.$ If $S_u$ is not empty
    then the node sends a point in $S_u$
    at maximum distance from the line segment between $p$ and $r$ 
    %having the largest $y$-coordinate and different from $p,\ r$ (if any)
    to the clique node holding $p$, from here on referred to as the
\emph{master node}.
%a master node (e.g., the node holding $p.$).
  \item
    If the master node has not received any point  satisfying the requirements
    from the previous step then it proclaims $p$ and $r$ to be
    vertices of the upper hull by sending this information
    to the nodes holding $p$ and/or $r,$ respectively.
    (In fact one of the vertices $p$ and $r$ has been marked as
    being on the upper hull earlier.)
    Next,
    it pops a call of $QuickUpperHull$ from the top of
    a stack of recursive calls
    held in a prefix of the clique nodes numbered $1,\ 2, ....$
    In case the stack is empty it terminates $QuickUpperHull(p_{min},p_{max}).$
  \item
    If the master node has received some points satisfying the requirements from Step 1
    then it determines a point $q$ at maximum
    distance from the line segment between $p$ and $r$
   % $y$-coordinate
    among them; see Fig. \ref{fig: pqr}.
    Next, it puts the call of $QuickUpperHull(q,r)$ on the top
    of the stack and then activates $QuickUpperHull(p,q).$
    \end{enumerate}

    The procedure $QuickLowerHull(p,r)$ is defined analogously.

    Each step of the procedure $QuickConvexHull(S),$
    but for the calls to\\
    $QuickUpperHull(p_{min},p_{max})$ and $QuickLowerHull(p_{min},p_{max})$,
    can be done in $O(1)$ rounds on
    the congested clique on $n$ nodes.  In particular, the sorting and
    the routing steps in $QuickConvexHull(S)$ can be done in $O(1)$ rounds by
Facts 1, 2.
    %Theorems 3.7 and 4.5 and their generalizations in \cite{L}.
    Similarly, each step of $QuickUpperHull(p,r),$ and symmetrically
    each step of $QuickLowerHull(p,r)$,
    but for recursive calls, can be done in $O(1)$ rounds.
    Since each non-leaf (in the recursion tree) call of $QuickUpperHull(p,r)$ and $QuickLowerHull(p,r)$
    results in a new vertex of the convex hull, their total number
    does not exceed $h.$
    Hence, we obtain the following theorem.

    \begin{theorem}
      Consider a congested $n$-clique network, where each node holds
      a batch of $n$ points in the Euclidean plane specified
      by $O(\log n)$-bit coordinates.
      Let $h$ be the number of vertices on the convex hull
      of the set $S$ of the $n^2$ points.
      The convex hull of $S$ can be computed
      by the procedure $QuickConvexHull(S)$ 
      in $O(h)$ rounds on the congested clique.
    \end{theorem}

    \section{An $O(\log n)$-round Algorithm for Convex Hull on Congested Clique}
\label{section: O(log_n)_rounds}

    Our refined algorithm for the convex hull of the input point set
    $S$ analogously as $QuickConvexHull(S)$
    described in Section \ref{section: quick}
    starts by sorting the
    points in $S$ by their $x$-coordinates and then splitting the
    sorted sequence of points in $S$ into an upper-hull subsequence
    and lower-hull subsequence. Next, it computes the upper hull of $S$
    and the lower hull of $S$ by calling the procedures
    $NewUpperHull(s)$ and $NewLowerHull(S),$ respectively. The
    procedure $NewUpperHull(S)$ lets each node $\ell$ construct the
    upper hull $H_{\ell}$ of its batch of at most $n$ points in
      the upper-hull subsequence locally. The crucial step of
      $NewUpperHull(S)$ is a parallel computation of bridges between
      all pairs $H_{\ell},$ $H_m$, $\ell \neq m,$ of the constructed
      upper hulls by parallel calls to the procedure $Bridge(H_{\ell},H_m)$.
      Based on the bridges between $H_{\ell}$
      and the other upper hulls $H_m,$ each node $\ell$ can determine
      which of the vertices of $H_{\ell}$ belong to the upper hull
      of $S$ (see Lemma 1). The procedure $Bridge$ has recursion
      depth $O(\log n)$ and the parallel implementation of the crucial
      step of $NewUpperHull(s)$ takes $O(\log n)$ rounds.
      The procedure $NewLowerHull(s)$ is defined symmetrically.
      Consequently, the refined algorithm for the convex hull
      of $S$ specified by the procedure $NewConvexHull(S)$
      can be implemented in $O(\log n)$ rounds.

      The procedure $NewConvexHull(S)$ is defined in exactly the same way
      as $QuickConvexHull(S)$,
except that the call $QuickUpperHull(p_{min},p_{max})$ in Step 6 is
replaced by the call $NewUpperHull(S)$ and the call
$QuickLowerHull(p_{min},p_{max})$ in Step 7 is replaced by the call
$NewLowerHull(S)$.

The strategy of $NewUpperHull(S)$ is to successively identify and mark points
in $S$ that cannot belong to the upper hull of $S$ as \emph{not qualifying},
and finally return the unmarked points.
\par
\vskip 5pt
\noindent
        {\bf procedure} $NewUpperHull(S)$
        \par
        \noindent
            {\em Input}: The upper-hull subsequence
            of the input point set $S$ held in consecutive
            nodes in batches of at most $n$ points.
            \par
            \noindent
                {\em Output}: The vertices of the upper hull of $S$
                held in clockwise order in consecutive nodes
                in batches of at most $n$ vertices.
           
  \begin{enumerate}
      \item
        Each node $\ell$ computes the upper hull $H_{\ell}$
        of its upper-hull subsequence locally.
      \item
        In parallel, for each pair $\ell, \ m$  of nodes,
        the procedure $Bridge(H_{\ell},H_m)$
        computing the bridge between $H_{\ell}$ and $H_m$ is called.
        (The procedure uses the two nodes in $O(\log n)$ rounds,
        exchanging at most two messages between the nodes
        in each of these rounds.)
      \item
        Each node $\ell$ checks if it has a single point $p$ not marked
        as not qualifying for the upper hull of $S$ such
        that there are bridges between $H_k$ and $H_{\ell}$
        and $H_{\ell}$ and $H_m$, where $k<\ell<m$, $p$
        is an endpoint of both bridges, and the angle formed
        by the two bridges is smaller than 180 degrees. If so,
        $p$ is also marked as
%not qualifying for the upper convex hull of $S.$ %%%
not qualifying for the upper hull of $S.$ %%%
      \item
        Each node $\ell$ 
prunes the set
of vertices of $H_l$, leaving only
those vertices that have not been marked
in the previous
steps (including calls to the procedure
$Bridge$) as not qualifying for the upper hull of $S$.
  \end{enumerate}

  The following lemmata enable the implementation of
  the $n^2$ calls to $Bridge(H_{\ell},H_m)$ in the second
    step of $NewUpperHull(S)$ in $O(\log n)$ rounds
    on the congested clique. 

\begin{figure}
\begin{center}
\includegraphics[scale=0.7]{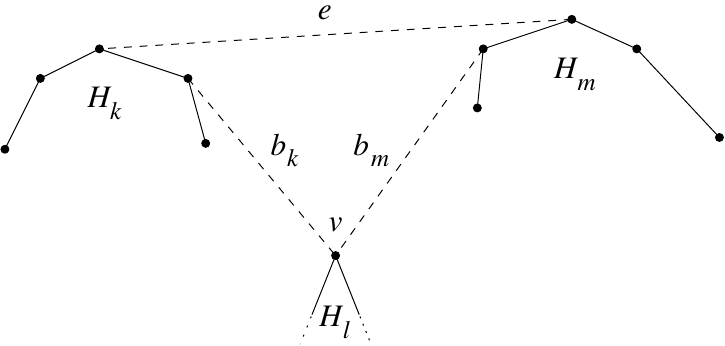}
\end{center}
\caption{The final case in the proof of Lemma 1.}
\label{fig: cccg4}
  \end{figure}

  \begin{lemma}
    %For $\ell \in [n],$
For any $\ell \in \{1,2,\dots,n\}$,
    let $H_{\ell}$ be the
    upper hull of the upper-hull subsequence of $S$
    assigned to the node $\ell.$
    A vertex $v$ of $H_{\ell}$ is not a vertex of the upper hull
    of $S$ if and only if it lies below a bridge
    between $H_{\ell}$ and $H_m,$ where $\ell\neq m,$
    or there are two bridges between $H_{\ell}$ and
    $H_k$, $H_m,$ respectively, where $k<\ell < m,$ such that
    they touch $v$ and form an angle of less than $180$ degrees
    at $v.$
  \end{lemma}\label{lem: first}
  \begin{proof}
    Clearly, if at least
    one of the two conditions on the right side of
    ``if and only if'' is satisfied
    then $v$ cannot be a vertex of the upper hull of $S.$
    Suppose that $v$ is not a vertex of the upper hull of $S.$
    Then, since it is a vertex of $H_l,$ there must be an edge $e$
    of the upper hull of $S$ connecting $H_k$ with $H_m$
    for some $k\le \ell \le m$, $k\neq m,$ that lies above $v.$
    We may assume without loss of generality that $v$ does not
    lie below any bridge between $H_{\ell}$ and $H_q,$ $\ell \neq q.$
    It follows that $k<\ell < m.$
  Let $b_k$ be the bridge between $H_k$ and $H_{\ell}$, and let
  $b_m$ be the bridge between $H_{\ell}$ and $H_m.$ It also follows that
  both $b_k$ and $b_m$ are placed below $e$ and the endpoint of $b_k$
  at $H_{\ell}$ is $v$ or a vertex of $H_{\ell}$ to the left
  of $v$ while the endpoint of $b_m$ at $H_{\ell}$ is $v$
  or a vertex to the right of $v.$ Let $C$ be the convex chain
  that is a part of $H_{\ell}$ between the endpoints of $b_k$
  and $b_m$ on $H_{\ell}.$ Suppose that $C$ includes at least one edge.
  The bridge $b_k$ has to form an angle
  not less than 180 degrees with the leftmost edge of $C$ and symmetrically
  the bridge $b_m$ has to form an angle not less than 180 degrees
  with the rightmost edge of $C.$ However, this is impossible because
  the bridges $b_k$ and $b_m$ are below the edge $e$ of the upper hull of $S$
  with endpoints on $H_k$ and $H_m$ so they form an angle less than 180 degrees.
  We conclude that $C$ consists solely of $v$ and consequently $v$
  is an endpoint of both $b_k$ and $b_m.$
See Fig.~\ref{fig: cccg4}. %%%
  \qed
  \end{proof}

\begin{figure}
%[h]
\begin{center}
\includegraphics[scale=0.7]{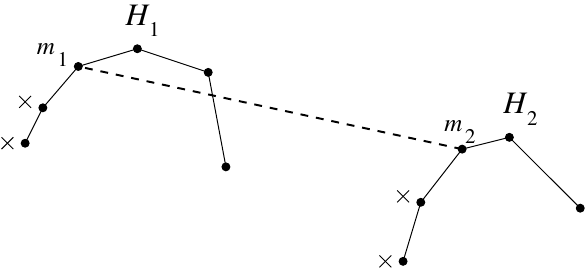}
\end{center}
\caption{An example of the segment connecting $m_1$ with $m_2$
in Lemma 2.}
\label{fig: nobridge}
  \end{figure}

  The following folklore lemma follows easily by  a standard case
  analysis (cf. \cite{KS86,OL81,P79}). See also Fig. \ref{fig: nobridge}.
  It implies that when computing the two endpoints of the bridge between two
upper hulls, one can eliminate at least a quarter of all the remaining
candidates after looking at six points only.
Hence, the recursive
   depth of the procedure $Bridge$ is $O(\log n).$

  \begin{lemma}\label{lem: nobridge}
    Let $S_1,\ S_2$ be two $n$-point sets in the Euclidean plane
    separated by a vertical line. Let $H_1,\ H_2$ be the upper hulls of
    $S_1,\ S_2,$ respectively. Suppose that each of $H_1$ and $H_2$ has
    at least three vertices.
    Next, let $m_1,\ m_2$ be the median vertices
    of $H_1,\ H_2,$ respectively. Suppose that the segment connecting $m_1$ with $m_2$
    is not the bridge between $H_1$ and $H_2$.
    %Then, the vertices on $H_1$ either to the left or to the right of $m_1,$
   % or the vertices on $H_2$ either to the left or to the right of $m_2$
    % cannot be an endpoint of the bridge between $H_1$ or $H_2.$
    Then depending on  $m_1,\ m_2$ and their neighbors on $H_1,\ H_2,$
    respectively, none of the vertices in at least one of the following four sets is an
endpoint of the bridge between $H_1$ and $H_2$:\\
(i)~the vertices on $H_1$ to the left of $m_1$; \\
(ii)~the vertices on $H_1$ to the right of $m_1$; \\
(iii)~the vertices on $H_2$ to the left of $m_2$; and\\
(iv)~the vertices on $H_2$ to the right of $m_2$.
\junk{   Then at least one of the following four cases holds:\\
(i)~none of the vertices on $H_1$ to the left of $m_1$,\\
(ii)~none of the vertices on $H_1$ to the right of $m_1$,\\
(iii)~none of the vertices on $H_2$ to the left of $m_2$, or\\
(iv)~none of the vertices on $H_2$ to the right of $m_2$\\
is an endpoint of the bridge between $H_1$ and $H_2$.}
  \end{lemma}
   \par
    \noindent
        {\bf procedure} $Bridge(H_{\ell}',H_m')$
        \par
        \noindent
            {\em Input}: A continuous fragment $H'_{\ell}$
            of the upper hull $H_{\ell}$ of
        the upper-hull subsequence assigned to a node $\ell$
        and a continuous fragment $H'_m$ of
        the upper hull $H_m$
        of the upper-hull subsequence assigned to the node $m.$
         \par
         \noindent
             {\em Output}: The bridge between $H'_{\ell}$ and $H'_m.$
             Moreover, all points in the upper-hull subsequence
             held in the nodes $\ell$ and $m$ placed
             under the bridge
             are marked as not qualifying for the convex hull of $S.$
             
        \begin{enumerate}
        \item
          If $H'_{\ell}$ or $H'_m$ has at most two vertices
then
%determine the bridge between $H'_{\ell}$ and $H'_m$, %%%
%e.g., by binary search. %%%
compute the bridge between $H'_{\ell}$ and $H'_m$ by
sending the at most two vertices from $\ell$ to $m$
or {\em vice versa}, computing the bridge locally
at $m$ or $\ell$, respectively, and sending back the bridge segment
from $\ell$ to $m$ or {\em vice versa}, respectively.
%binary search. %%%
Next, mark
          all the
          points in the upper-hull subsequence between the endpoints
          of the found bridge
        that are assigned to the nodes $\ell$ or $m$ as not qualifying
        for vertices of
the upper hull of $S$ and stop.
      \item
%        Determine %%%
        Find %%%
a median $m_1$ of $H'_{\ell}$ and a median $m_2$ of $H'_m$.
      \item
If the straight line passing through $m_1$ and $m_2$ is a
        supporting line for both $H'_{\ell}$ and $H'_m$ then mark all the
        points in the upper-hull subsequence between $m_1$ and $m_2$
        that are assigned to the nodes $\ell$ or $m$ as not qualifying
        for vertices of the upper hull of $S$ and stop.

      \item 
        Otherwise, $\ell$ and $m$ inform each other about the neighbors
        of $m_1$ on $H'_{\ell}$ and the neighbors of $m_2$ on  $H'_m,$ respectively.
        Then, $Bridge(H''_{\ell},H''_m)$ is called, where
         either $H'_{\ell}=H''_{\ell}$ and $H''_m$
         is obtained from $H'_m$ by removing vertices on
%one of the sides %%%
         %one side %%%
         the appropriate side
         of the median of $H'_m$ or {\em vice versa},
         according to Lemma 2.
    \end{enumerate}

    The procedure $NewLowerHull(H'_{\ell},H'_m)$ is defined analogously.

    As in the procedure $QuickConvexHull(S),$
    each step of $NewConvexHull(S),$ but for the
    calls to $NewUpperHull(S)$ and
    $NewLowerHull(S)$, can be done in $O(1)$ rounds
    on the congested clique by \cite{L}.
    Furthermore, the first, next to the last,  and last steps of $NewUpperHull(S)$
    require $O(1)$ rounds. By Lemma 2, the recursion depth
    of the procedure $Bridge$
    is logarithmic in $n.$
    The crucial observation is now
    that consequently the nodes $\ell$ and $m$ need to exchange
    $O(\log n)$ messages in order to implement $Bridge(H_{\ell},H_m)$.
    In particular, they need to inform each other about the
    current medians and their neighbors
    on $H'_{\ell}$ or $H'_m$, respectively.
    Also, in case $H'_{\ell}$ or $H'_m$ contains
    at most two vertices, the node $\ell$ or $m$ needs to
    inform the other node about the situation and about those at most
    two vertices. In consequence, by Lemma 2,
    these two nodes can implement  $Bridge(H_{\ell},H_m)$
    by sending a single message to each other in each
    round in a sequence of $O(\log n)$ consecutive rounds.
    It follows that all the $n^2$ calls of $Bridge(H_{\ell},H_m)$
    can be implemented in parallel in $O(\log n)$ rounds.
    Note that in each of the $O(\log n)$ rounds, each
    clique node sends at most one message to each other
%    clique node,so in total, %%%
    clique node, so in total, %%%
    each node sends at most $n-1$
    messages to the other nodes in each of these rounds.
    It follows that $NewUpperHull(S)$ and symmetrically
    $NewLowerHull(S)$ can be implemented in $O(\log n)$
    rounds on the congested clique.
    We conclude that $NewConvexHull(S)$
    can be done in $O(\log n)$
    rounds on the congested clique.
      
\begin{theorem}
      Consider a congested $n$-clique network, where each node holds
      a batch of $n$ points in the Euclidean plane specified by
      $O(\log n)$-bit coordinates.
      The convex hull of the set $S$
      of the  $n^2$ input points can be computed
      by the procedure
      $NewConvexHull(S)$
      in $O(\log n)$ rounds on the congested clique.
    \end{theorem}

\section{Point Set Triangulation in $O(\log^2 n)$ Rounds on Congested Clique}
Our method of triangulating a set of $n^2$ points in the congested $n$-clique
model initially resembles that of constructing the convex hull of
the points. That is, first the input point set is sorted by $x$-coordinates.
Then, each node triangulates its sorted batch of $n$ points locally.
Next, the triangulations are pairwise merged and extended to triangulations
of doubled point sets by using the procedure $Merge$
in parallel in $O(\log n)$ phases. In the general case, the procedure $Merge$
calls the procedure $Triangulate$ in order to triangulate the area
between the sides of the convex hulls of the two input triangulations,
facing each other.

\begin{figure}
%[h]
\begin{center}
\includegraphics[scale=0.7]{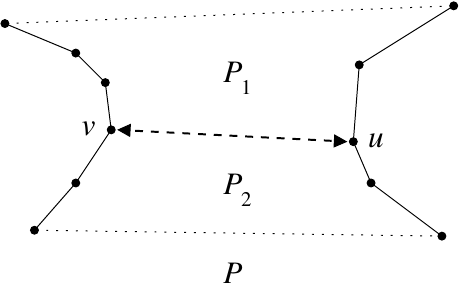}
\end{center}
\caption{An example of the partition of the polygon $P$
into the subpolygons $P_1,\ P_2$ in the procedure $Triangulate$.}
\label{fig: triangulate2}
  \end{figure}

The main idea of the procedure $Triangulate$
is to pick a median vertex $v$ on the longer of the convex hulls sides
and send its coordinates and the coordinates
of its neighbors to the nodes holding the facing side
of the other hull. The latter nodes send back candidates (if any)
for a mate $u$ of the median vertex $v$ such that the segment between
$v$ and $u$ can be an edge of
a triangulation extending the existing partial triangulation. The segment is used
to split the area to triangulate into two that are triangulated
by two recursive calls of $Triangulate$ in parallel.
See Fig. \ref{fig: triangulate2}.
Before the recursive calls the edges of the two polygons surrounding
the two areas are moved to new node destinations so each of the
polygons is held by a sequence of consecutive clique nodes.
This is done by a global routing in $O(1)$ rounds serving
all parallel calls of $Triangulate$ on a given recursion level,
for a given phase of $Merge$ (its first argument).

Since the recursion depth
$Triangulate$ is $O(\log n)$ and $Merge$ is run in $O(\log n)$ phases,
the total number of required rounds becomes $O(\log^2 n).$

To simplify the presentation, we shall assume that the size $n$ of the clique network
is a power of $2.$
\newpage
%\par
 %   \vskip 5pt
    \noindent
    {\bf procedure} $Triangulation(S)$

    \begin{enumerate}
    \item Sort the points in $S$ by their $x$-coordinates so
      each node receives a subsequence consisting of $n$ consecutive
      points in $S$, in the sorted order.

 \item
    Each node  sends the first point and the last point in its subsequence
    to the other nodes.

  \item
    Each node $q$ constructs a triangulation $T_{q,q}$ of the points in its
    sorted subsequence locally.
  \item
    For $1\le p< q\le n,$ $T_{p,q}$ will denote the already computed
    triangulation of the points in the sorted subsequence held
    in the nodes $p$ through $q.$
    For $i=0,\dots,\log n -1,$ in parallel,
    %for $j=1, 1+2^{i+1}, 1+22^{i+1},1+32^{i+1},...$
    for $j=1, 1 + 2^{i+1}, 1 + 2 \cdot 2^{i+1}, 1 + 3 \cdot 2^{i+1}, \dots$
    the union of the triangulations $T_{j,j+2^i-1}$ and
    $T_{j+2^i,j+2^{i+1}-1}$ is transformed
    to a triangulation $T_{j, j+2^{i+1}-1}$
    of the sorted subsequence held 
    in the nodes $j$ through $j+2^{i+1}-1$ by calling the procedure
    $Merge(i,j)$.
    \end{enumerate}
    \par
    \noindent
        {\bf procedure} $Merge(i,j)$
        \par
        \noindent
            {\em Input}: A triangulation $T_{j,j+2^i-1}$
            of the subsequence held
            in the nodes $j$ through $j+2^i-1$ 
            and a triangulation $T_{j+2^i,j+2^{i+1}-1}$ of the subsequence
            held in the nodes $j+2^i$ through
            $j+2^{i+1}-1,$.
            \par
            \noindent
                {\em Output}: A triangulation $T_{j, j+2^{i+1}-1}$
                %$T_{j, j+2{j+1}-1}$
                of the subsequence held
                in the nodes $j$ through $j+2^{j+1}-1$.

                \begin{enumerate}
                \item
                  Compute the bridges between the convex hulls of $T_{j,j+2^i-1}$ and $T_{j+2^i, j+2^{i+1}-1}$.
                  Determine the polygon $P$ formed by the bridges between
                  the convex hulls of $T_{j,j+2^i-1}$ and $T_{j+2^i,j+2^{i+1}-1}$,
                  the right side of the convex hull of $T_{j,j+2^i-1}$, and
                  the left side of the convex hull of $T_{j+2^i,j+2^{i+1}-1}$
                  between the bridges.
                \item
                  $Triangulate(P,j,j+2^{i+1}-1)$
                \end{enumerate}
                \par
    \noindent
        {\bf procedure} $Triangulate(P,p,q)$
        \par
        \noindent
            {\em Input}: A simple polygon $P$ composed of two convex chains
            facing each other on opposite sides of a vertical line
            and two edges crossing the line, held in
%the nodes $p$ through $q,$ $p<q.$ %%%
nodes~$p$ through~$q,$ with $p<q.$ %%%
            \par
            \noindent
           {\em Output}: A triangulation of $P$ held in
%the nodes $p$ through $q.$ %%%
nodes~$p$ through~$q.$ %%%

           \begin{enumerate}
       \item
         If $p=q$ then the $p$ node triangulates $P$ locally and terminates
         the call of the procedure.
      \item
        The nodes $p$ through $q$ determine the lengths of the convex chains
        on the border of $P$ and the node holding the median vertex $v$ of the longest
        chain (in case of ties, the left chain) sends the coordinates
        of $v$ and the adjacent vertices on the chain
        to the other nodes $p$ through $q.$
      \item
        %The nodes holding the vertices of the chain opposite to that
        The nodes holding vertices of the convex chain that is
        opposite to the convex chain containing $v$ determine if they
        hold vertices $u$ that could be connected by a segment with $v$
        within $P.$ They verify if the segment $(v,u)$
        is within the intersection of the union of
        the half-planes on the side of $P$ induced by the edges adjacent to $v$ with
        the union of the half-planes on the side of $P$  induced by the edges adjacent to $u$.
        If so, they send one such a candidate vertex $u$
        to the node holding $v.$
        \item
        The node holding $v$ selects one of the received candidate vertices $u$ as
        the mate and sends its coordinates to the other nodes $p$ through $q.$
      \item
        The nodes $p$ through $q$ split the polygon $P$ into two subpolygons
        $P_1$ and $P_2$ by the edge $(v,u)$ and by exchanging messages in $O(1)$ rounds compute
        the new destinations  for the edges of the polygons $P_1$ and $P_2$
        so $P_1$ can be held  in nodes $p$ through $r_1$ and $P_2$ in the
        nodes $r_2$ through $q$, where $p\le r_1\le r_2\le q$ and $r_1=r_2$
        or $r_2=r_1+1.$
      \item
        A synchronized global routing in $O(1)$ rounds corresponding to the current
        phase of the calls to the procedure $Merge$ (given by its first argument)  and all parallel calls
        of the procedure $Triangulate$ on the same recursion level
        is implemented by using Fact 1. In particular, the edges of $P_1$ and $P_2$ are moved
        to the new consecutive destinations among
%the %%%
nodes $p$ through $q.$ %%%
      \item
        In parallel, $Triangulate(P_1,p,r_1)$ and $Triangulate(P_2,r_2,q)$ are performed.
           \end{enumerate}

           At the beginning, we have outlined our triangulation method, in particular the procedures
           forming it, in a top-down fashion.
% We complement this outline with a more bottom-up analysis. %%%
We now complement this outline with a bottom-up analysis.
           All steps of the procedure $Triangulate(P,p,q)$ but for
           the recursive calls in the last step and the next to the last step
           can be implemented in $O(1)$ rounds, using only the nodes $p$ through $q.$ The next to the last step
           is a part of the global routing.
           It serves all calls of the procedure $Triangulate$
           on the same recursion level for a given phase of the parallel calls
           of procedure $Merge(i,\ )$, i.e.,  for given $i.$ Since each node is involved in
           at most two of the aforementioned calls of $Triangulate$
           that cannot be handled locally, the
           global routing, implementing the next to the last step of $Triangulate$,
           requires $O(1)$ rounds. Since the recursion depth of $Triangulate$ is $O(\log n)$,
%           we conclude that %%%
$Triangulate$ takes $O(\log n)$ rounds.
           The first step of the procedure $Merge(i,j)$, i.e., constructing the
           bridges between the convex hulls,
           can be implemented in $O(\log n)$ rounds by using
           %the procedure $Bridge$ from
           the convex hull algorithm from
% the preceding section %%%
Section~\ref{section: O(log_n)_rounds}
on
%the %%%
nodes $j$ through $j+2^{i+1}-1.$
%           The second step can be easily %%%
           The second step can easily be
implemented in $O(1)$ rounds using the aforementioned
           nodes.
Finally, the call
%of %%%
to %%%
% the procedure %%%
$Triangulate$
in the last step of $Merge$
           requires $O(\log n)$ rounds by our analysis of this procedure.
           Again, it can be done by
%the %%%
nodes $j$ through $j+2^{i+1}-1$ but for
           the last steps of calls to $Triangulate$ that are served by the discussed synchronized
           global routing in $O(1)$ rounds. We conclude that $Merge(i,j)$ can be implemented
           in $O(\log n)$ rounds.
           Finally, all steps in
%the procedure %%%
$Triangulation(S)$
%           but for the step %%%
except the one %%%
involving parallel calls to $Merge(i,j)$
           in $O(\log n)$ phases can be done in $O(1)$ rounds.
           For a given phase, i.e., given $i,$ each node is involved
           in $O(1)$ calls of $Merge(i,j)$ but for the next to the last steps
           in $Triangulate$ that for a given recursion level of
           $Triangulate$ are implemented by the joint global routing
           in $O(1)$ rounds. It follows from
           our analysis of $Merge(i,j)$
           and $i=O(\log n)$ that $Triangulate(S)$ can
           be implemented
%in $O(\log^2 n)$ rounds on the congested clique. %%%
in $O(\log^2 n)$ rounds.

     \begin{theorem}
      Consider a congested $n$-clique network, where each node holds
      a batch of $n$ points in the Euclidean plane specified
      by $O(\log n)$-bit coordinates.
      A triangulation of the set $S$ of the $n^2$ input  points can be computed
      by the  procedure $Triangulation(S)$
      in $O(\log^2 n)$ rounds on the congested clique.
     \end{theorem}
     
    \section{On the Construction of Voronoi Diagram
     on Congested Clique}
     
    The primary difficulty in the design of efficient parallel algorithms
     for the Voronoi diagram of a planar point set
     using a divide-and-conquer approach 
     is the efficient parallel merging of Voronoi diagrams.
     In \cite{ACG85}, Aggarwal et al. presented a very involved
     $O(\log n)$-time PRAM method for the parallel merging.
     As a result, they obtained an $O(\log ^2 n)$-time CREW PRAM algorithm
     for the Voronoi diagram. Their work and later improved
     PRAM algorithms for the Voronoi diagram
     \cite{CGO,VV} suggest that this problem should be solvable in $(\log n)^{O(1)}$
    % $(\log n)^{O(1)}$
     rounds on the congested clique.
     %In the full version of this paper, we will show

     When the points with $O(\log n)$-bit coordinates are drawn
     uniformly at random from a unit square or circle then the
     expected number of required rounds to compute the Voronoi diagram
     or the dual Delaunay
     triangulation
     on the congested clique becomes $O(1)$ (cf. \cite{LKL,VV}).
     To demonstrate this
     %fact one can use the probabilistic analyses given in the two
     %papers on $O(\log n)$ expected-time PRAM algorithms for the
     %Voronoi diagram of a set of $n$ points drawn from a unit square
     %or circle uniformly at random \cite{LKL,VV}.
     %The authors
     %are currently working on a round-efficient (in expectation)
    % algorithm for the Voronoi diagram in this case on the congested clique.
     we need to recall the Chernoff bounds.

     \begin{fact} \label{fact: chernoff}(multiplicative Chernoff lower bound)
  Suppose $X_1, ..., X_n$ are independent random variables
  taking values in $\{0, 1\}.$
  Let X denote their sum and let $\mu= E[X]$
  denote the sum's expected value. Then, for any $\delta \in [0,1],$
  $Prob(X\le (1-\delta)\mu)\le e^{-\frac {\delta^2 \mu}2}$ holds.
  Similarly, for any $\delta \ge 0,$
  $Prob(X\ge (1+\delta)\mu)\le e^{-\frac {\delta^2 \mu}{2+\delta}}$ holds.  
  \end{fact}

     We shall say that an event dependent on $n^2$ input points in the
     plane holds with high probability (w.h.p.) if its probability is
     at least $1-\frac 1 {n^{\alpha}}$ asymptotically, (i.e., there is
     an integer $n_0$ such that for all $n\ge n_0,$ the probability is
     at least $1-\frac 1 {n^{\alpha}}$), where $\alpha $ is a
     constant not less than $2$.

  \begin{theorem}\label{theo: vor}
       The Voronoi diagram of $n^2$ points with $O(\log n)$-bit coordinates
       drawn uniformly at random from a unit square
       in the Euclidean plane can be computed within the square w.h.p.
       in $O(1)$ rounds on the congested clique.
     \end{theorem}

     \begin{proof}
       Consider an arbitrary square $R$ of size $\frac 1 {\sqrt n}  \times
       \frac 1 {\sqrt n} $ within the unit square. Next,
       consider a sequence of $n^2$ uniform random draws of points
       with $O(\log n)$-bit coordinates from the unit square. Call a
       draw in the sequence a success if a point within $R$ is
       drawn. The expected number of successes is $n.$
       %Hence, for
       %any fixed $\alpha > 2,$ it follows from the Chernoff bounds that
      % $\Theta(n)$ of the drawn points are within $R$ with
       %probability at least $1- \frac 1{n^{\alpha}}.$
       Hence, it follows by selecting $\delta = \sqrt{\frac{6 \ln n}{n}}$ in the
Chernoff bounds that $\Theta(n)$ of the drawn points are within $R$ with
probability at least $1 - \frac{1}{n^{3}}$.

         Partition the unit square into $n$
         rectilinear squares of size $\frac 1 {\sqrt n}  \times
       \frac 1 {\sqrt n}$.

\begin{figure}
%[h]
\begin{center}
\includegraphics[scale=0.7]{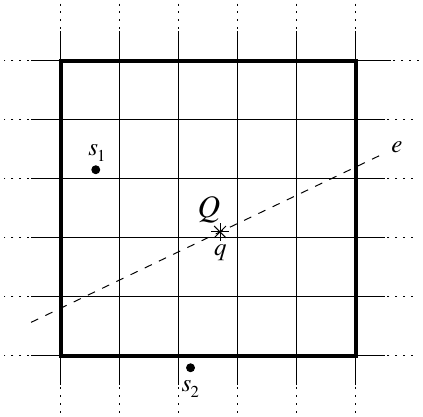}
\end{center}
\caption{An example of the configuration in  the proof of Theorem \ref{theo: vor}.}
\label{fig: theo}
  \end{figure}
       
Let $S$ be the set of $n^2$ points drawn from the unit
         square.  Consider the Voronoi diagram of $S$ within the unit
         square.  Let $e$ be an edge of the Voronoi diagram. The edge
         $e$ has to be a part of the bisector of some couple of points
         $s_1$ and $s_2$ in $S.$ Consider an arbitrary point $q$ on
         $e$ and the rectilinear square $Q$ in the aforementioned
         partition that contains it.
Suppose that $s_1$ or $s_2$ lies outside the rectilinear area formed by $Q$ and
the two layers of squares around $Q$ in the partition, i.e., consisting of at
most $1+8+16=25$ squares including $Q$.  See Fig. \ref{fig: theo}.  Without loss of
generality, let $s_2$ be such a point.  Then the distance between $q$ and $s_2$
is at least $2 \cdot \frac{1}{\sqrt n}$, while the distance between $q$ and
every point inside~$Q$ is at most $\sqrt{2} \cdot \frac{1}{\sqrt n}$.  We
obtain a contradiction w.h.p. because $Q$ contains $\Omega(n)$ points from $S$
w.h.p. and $q$ is closer to each of these points than to $s_2$.
\junk{Suppose that $s_1$ or $s_2$ is
         outside $Q$ as well as outside the two layers of the rectilinear
         squares around $Q$
         in the partition, together with $Q$ at most
         $1+8+16=25$ squares. We obtain a contradiction
         w.h.p. since $Q$ contains $\Omega (n)$ points in $S$ w.h.p. They
         are closer to $q$ than the aforementioned point outside the
         at most $25$ squares.}
         It follows that to compute
         the Voronoi diagram of $S$ within a square in the partition
         w.h.p. one needs solely to know the points in $S$ located
         in $Q$ and the at most $24$ squares around the square.  

         We can assign to each of the
         squares in the partition a distinct clique node and deliver to each node the points from its square in $O(1)$ rounds w.h.p.
         by using the sorting and routing $O(1)$-round algorithms
         from Facts 1, 2.
        % \cite{L}.
         Then, additionally we need to deliver
         to each node the points in $S$ located in the at most $24$
         squares around its square.
         By using again the routing
         algorithms from Facts 1, 2,
         %\cite{L},
         this can be achieved in
         $O(1)$ rounds w.h.p.
         (Note that the total number of points
         that need to be delivered to each node
         is $O(n)$ w.h.p. since each of the squares
         contains $O(n)$ points w.h.p.)
        % Finally, each node computes
        % locally the Voronoi diagram of the w.h.p. $O(n)$ many
        % points it received
Finally, each node applies any sequential Voronoi diagram
         algorithm (e.g., \cite{F87}) to locally compute the Voronoi
         diagram of the w.h.p. $O(n)$ many points it received and then
         it determines its intersection with the square assigned to
         the node.
         %It follows by enhancing the constant $\alpha$ in
         %the analysis of the distribution of drawn points in the
         %arbitrary square $R$ and the union bound that all nodes
         %compute the Voronoi diagram within their squares w.h.p.  \qed
         This shows that each node computes its local Voronoi diagram correctly w.h.p.,
and according to the first paragraph above, this probability is at least
$1 - \frac{1}{n^{3}}$.
Now, the union bound implies that the probability that all nodes compute their
local Voronoi diagrams correctly is at least $1 - \frac{1}{n^{2}}$.
     \end{proof}

\section{Concluding Remarks}
We have provided the first non-trivial, polylogarithmic upper bounds
on the number of rounds required to construct the convex hull and a
triangulation of a set of $n^2$ points in the plane with $O(\log
n)$-bit coordinates in the model of congested clique.  As for the
construction of the Voronoi diagram of the point set, we have shown an
$O(1)$ upper bound on the number of rounds under the assumption that
the points are drawn uniformly at random from a unit square. The major
open problem is the derivation of a non-trivial upper bound on the
number of rounds sufficient to construct the Voronoi diagram when the
points are not necessarily randomly distributed.  This seems to be
possible but it might require a substantial effort; see the discussion
in the preceding section.  An interesting question is also if a simple
polygon on $n^2$ vertices with $O(\log n)$-bit coordinates can be
triangulated using a substantially smaller number of rounds than that
needed to triangulate a set of $n^2$ points in the plane with $O(\log
n)$ bit coordinates in the congested $n$-clique model.
     
     \section*{Acknowledgments}
  This research was partially supported by Swedish Research Council grants
621-2017-03750 and 2018-04001,
 and JSPS KAKENHI JP20H05964.
\junk{J.J. was partially supported by JSPS KAKENHI JP20H05964
and Swedish Research Council grant 621-2017-03750.
JJ was partially supported by
the Algorithmic Foundations for Social Advancement Project's
Grant-in-Aid for Transformative Research Areas, MEXT, Japan.}

%---------------------------- Bibliography -------------------------------

% Please add the contents of the .bbl file that you generate,  or add bibitem entries manually if you like.
% The entries should be in alphabetical order
%\newpage
 \small
\bibliographystyle{abbrv}

\end{document}